\documentclass[aps,prl,twocolumn,showpacs,nofootinbib]{revtex4}

\usepackage{amsmath}
\usepackage{graphicx}
\usepackage{calc}
\usepackage{bm}

\usepackage[T2A]{fontenc}
\usepackage[cp1251]{inputenc}

\begin{document}

\title{Two-dimensional system of strongly interacting electrons in silicon (100) structures}

\author{V.~T. Dolgopolov}
\affiliation{Institute of Solid State Physics, Russian Academy of Sciences, Chernogolovka, Moscow District, 142432 Russian Federation
}

\begin{abstract}
Studies of different experimental groups that explore the properties of a two-dimensional electron gas in silicon semiconductor systems ((100) Si metal-oxide-semiconductor field-effect transistors (MOSFETs) and (100) SiGe/Si/SiGe quantum wells) in the vicinity of the metal-insulator transition are described and critically analyzed. Results are identified that are common to all research: (i) the effective mass of electrons measured at the Fermi level in the metallic regime increases as the electron density decreases and, if extrapolated, tends to diverge; (ii) the behavior of the energy-averaged mass in the metallic region is quite different in the two systems: in Si-MOSFETs, it also exhibits a tendency to diverge, while in the SiGe/Si/SiGe quantum wells it saturates in the limit of low electron densities; (iii) there is a small number (depending on the sample quality) of localized electrons in the metallic phase; (iv) the properties that the electron system exhibits in the insulating phase in the vicinity of the metal-insulator transition are typical of amorphous media with a strong coupling between particles.

\end{abstract}
\maketitle

\textbf{Contents}
\par\bigskip
{\bf 1. Introduction}

{\bf 2. Electrons in (100) Si-MOSFETs}

\hspace{4mm} 2.1	Metal-insulator transition in the absence of a magnetic field

\hspace{4mm} 2.2.  Influence of a magnetic field on the metal-insulator transition

\hspace{4mm} 2.3. Electron properties in the depth of the Fermi distribution

\hspace{4mm} 2.4. Electron properties at the Fermi level

\hspace{4mm} 2.5. Intermediate conclusions

\hspace{4mm} 2.6.    Electrons in the insulator

\hspace{4mm} 2.7. Additional intermediate conclusions 

{\bf 3.  Electrons in SiGe/Si/SiGe quantum wells}

\hspace{4mm} 3.1. Advantages and disadvantages of the structures

\hspace{4mm} 3.2. The tendency of a flat band to appear at the Fermi level 

{\bf 4.   Conclusions}

\hspace{4mm} \textbf{References}

\par\bigskip
\par\bigskip

{\bf 1.	Introduction}

\par\bigskip

The review is devoted to a brief description of state-of-the-art experimental studies of strongly correlated two-dimensional electron systems based on silicon semiconductor structures. An electron system is called strongly correlated if the characteristic Coulomb interaction energy between electrons greatly exceeds their kinetic (Fermi) energy  $\varepsilon_F$. Because the former is inversely proportional to the mean distance  between  electrons  (\textit{i.e.}, $\sqrt (n_s^{-1})$, where $n_s$ is the density of carriers, and $\varepsilon_F \propto n_s$, a strong coupling corresponds to low electron densities. In most experiments discussed below, the electron gas is assumed to be degenerate, \textit{i.e.}, $\varepsilon_F \gg kT$, where  $k$ is the Boltzmann constant and  $T$ is the temperature.

The electron-electron interaction strength is usually characterized by a parameter $r_s$ equal to the ratio of the Wigner-Seitz  radius 
$(\pi n_s)^{-1/2}$ to  the  Bohr  radius  of  the electron, $a_B= \frac{\kappa \hbar^2}{me^2}$, where   $\kappa$ is  the  dielectric  constant and $m$ and  $e$ are the electron mass and charge. In the simplest case of a single-valley electron system (which does not include the electron systems considered below), $r_s$ is equal to the ratio of the  characteristic potential interaction energy to the Fermi energy.

\begin{figure}
\scalebox{0.7}{\includegraphics[width=\columnwidth]{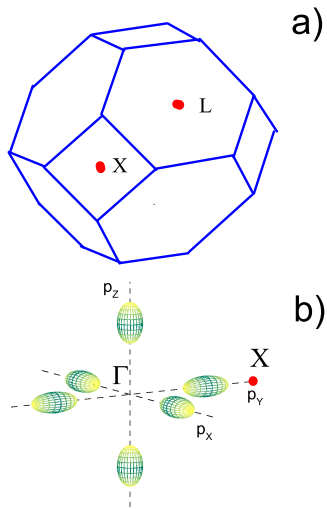}}
\caption{(a) Wigner-Seitz cell for electrons in silicon. (b) Image of
the equipotential surfaces near the minima of the conduction band.
}
\label{Figbrill}
\end{figure}

For the readers who are not familiar with the properties of silicon-based two-dimensional electron structures, we very schematically present some information. In the momentum space, the Wigner-Seitz cell for silicon is a truncated octahedron (Fig.\ref{Figbrill}(a)). The center of the Wigner-Seitz cell is denoted by $\Gamma$, the center of squares by X, and the center of hexagons by L. The minima of the conduction band are  located on  straight lines connecting points $\Gamma$ and X.	There are six such points altogether. The equipotential surfaces in the momentum space are shown schematically in Fig.\ref{Figbrill}(b). They are ellipsoids with a mass of  $0.98\, m_0$ along the major axis (where $m_0$ is the free electron mass) and  $m_b=0.19\,m_0$ for the momentum in the perpendicular direction.

If potential barriers are introduced making the electron motion in the $z$-direction restricted (for the (100) sample orientation), discrete levels appear in such a quantum well, the lowest of them being determined by ellipsoids with the major axis directed along $z$ (with the ``heavy'' mass corresponding to the motion along $z$). There are two such ellipsoids, and therefore the spectrum of the form

\begin{equation}
\varepsilon (p)= \varepsilon_0 +p^2/2m_b,
\label{eqsp}
\end{equation} 
where $\varepsilon_0$ is the energy of the lowest level in the quantum well and $p$ is the momentum in the (100) plane, has the so-called ``valley'' degeneracy in addition to spin degeneracy. Strictly speaking, the valley degeneracy is lifted in an asymmetric potential well; however, we ignore this small splitting in what follows.

To fill a potential well with electrons, the Fermi energy  $\varepsilon_F$ must  exceed  $\varepsilon_0$. With  this  condition  fulfilled,  the wave functions of the electrons	 freely moving along the quantum well spread  in  the  $z$  direction  by  
 $30$~\AA~in Si-MOSFETs and by  $150$~\AA~in SiGe/Si/SiGe quantum wells considered in Section 3. 

Two electron systems considered here have another substantial difference. In silicon MOSFETs, electrons are localized at the interface between silicon, with the  dielectric constant $\kappa_{Si}$, and silicon dioxide  ($\kappa_{SiO_2}$). Therefore, the interaction  between  electrons  is  determined  by  the  average dielectric constant $(\kappa_{Si}+\kappa_{SiO_2})/2 \simeq 7.7$.
In the SiGe/Si/SiGe quantum well, the interaction is determined by the dielectric constant  $\kappa_{SiGe}$ close to  $\kappa_{Si}$. Therefore, to achieve the same interaction strength in SiGe/Si/SiGe quantum wells, it is necessary to reduce the electron density by at least a factor of 2.5.

The  carrier density is controlled by a metallic electrode (gate) deposited on the dielectric separating a two-dimensional electron layer from the gate. The electron density depends linearly on the potential difference between the gate and the electron layer. We show below that the gate allows obtaining information on many properties of the electron system.

\par\bigskip  
    {\bf 2.	Electrons in (100) Si-MOSFETs}
\par\bigskip    

{\bf   2.1. Metal-insulator transition
in the absence of a magnetic field}\\

Two-dimensional electron systems at liquid helium temperatures can have high conductivity, considerably (by more than two orders of magnitude) exceeding  $\sigma_0 = e^2/ h $ (where $h$ is Planck constant).   However, at low electron densities, real electron systems exhibit a low conductivity $\sigma\ll \sigma_0$, with the activation temperature dependence typical of the insulator.

Starting with paper  \cite{band}, which is based on the scaling hypothesis, it has become commonly accepted that a metallic phase in a two-dimensional  electron system  (even with an arbitrarily weak disorder) is impossible in the sense that such a system with infinite dimensions at zero temperature would inevitably become  an  insulator.  In  this  case,  the  metal-insulator 
transition  (MIT) in a two-dimensional system turned  out to be impossible, and the experimentally observed transition was
called  ``apparent''.  The  term  is  somewhat  misleading because ``apparent'' should rather refer to the insulator expected at unrealistically low temperatures and fantastically huge dimensions of the sample.

\begin{figure}
\scalebox{0.8}{\includegraphics[width=\columnwidth]{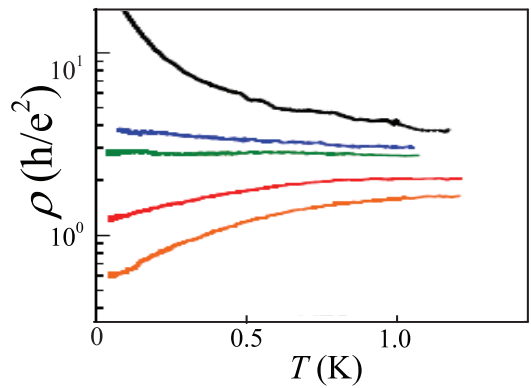}}
\caption{Temperature dependences of the resistance of an Si-MOSFET sample in the absence of a magnetic field (a) at electron densities (from top down) $6.85, 7.17, 7.25, 7.57, 7.85\times10^{10}$ cm$^{-2}$. From Ref.\cite{kl}. }
\label{Figsc}
\end{figure}

A revolutionary role was played by experiments  \cite{gesha,gesha1}, in which the unusual temperature behavior of the resistance of highly mobile electrons in Si-MOSFETs was observed in the vicinity of the transition: the resistance as a function of temperature changed its behavior typical of a metal to that typical of an insulator (see Fig.\ref{Figsc}). In fact, this effect was observed even earlier in \cite{zavar}, but remained unnoticed.

It was found in  Refs.~\cite{gesha,gesha1} that $\rho(T)$ dependences in both the metallic and insulating regimes can be scaled onto two universal curves (here $\rho$ is the resistivity). These two curves are separated by a temperature-independent line corresponding to the density $n_c=7.25\times10^{ 10}$~cm$^{-2}$ (Fig.\ref{Figsc}). By extrapolating this line to zero temperature, we see that the MIT in the electron system under study is indeed possible even at a zero temperature in samples with infinite dimensions, which obviously contradicts the conclusions of Ref.\cite{band}. Since at electron densities of the order of  $10^{11}$~cm$^{-2}$ the kinetic (Fermi) energy in the electron gas under study is an order of magnitude lower than the characteristic electron-electron interaction energy, this contradiction was interpreted as the result of strong coupling between electrons. It seemed that the one-parameter scaling remained valid with the scaling function modified by the interactions. The conviction that the one-parameter scaling remains universal has led to paradoxical conclusions \cite{popovic,popovic1} that a two-dimensional electron system remains metallic even for resistances of the order of   $3\times10^7$~Ohm per square and a positive temperature derivative of the conductivity, and that the critical electron density $n_c$ can decrease with increasing disorder.
 
At the same time, more realistic renormalization group calculations have been made taking renormalization with the increasing coupling strength and disorder into account \cite{fink4,fink, fink1,fink2,fink3}. Based on such calculations, a conclusion has been made that the phase transition observed in the most perfect Si-MOSFETs is indeed a quantum phase transition \cite{fink1,fink2}. Experimental data were consistent with the theory in the metallic regime \cite{fink3,kn2,fink5}. Paper \cite{fink5} is of interest because it describes the experimental temperature dependence of the resistance in a considerably broader range than the theory of small corrections does \cite{dolg,al}.

\begin{figure}
\scalebox{0.8}{\includegraphics[width=\columnwidth]{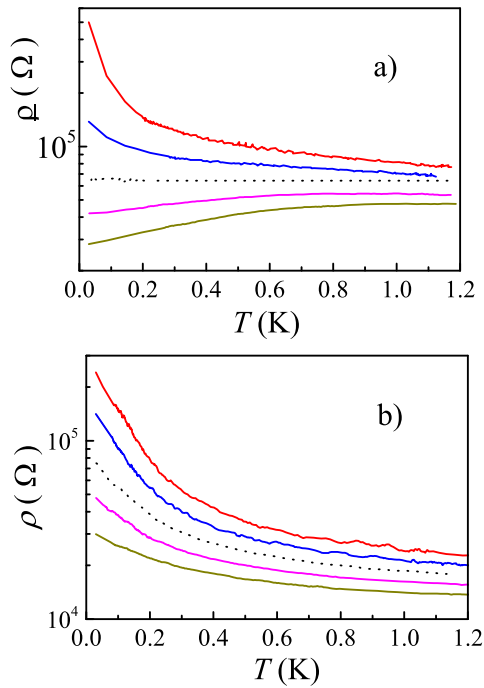}}
\caption{Temperature dependences of the resistance of a Si-MOSFET sample in the absence of a magnetic field at electron densities (from top down)  $7.65, 7.8, 7.95, 8.1, 8.25\times10^{10}$ cm$^{-2}$ (a) and in the 4~T magnetic field parallel to the interface and (b) at electron densities   $1.095, 1.125, 1.155, 1.185, 1.215\times 10^{11}$ cm$^{-2}$. From \cite{sh11}.}
\label{Fig11sc}
\end{figure}

The subject of the MIT in two dimensions (in particular, in highly mobile MOSFETs) has been considered in many reviews \cite{sar,sar1,pud04,sh2,sar2,popovic,sh1,dolgop}. We do not repeat their content but only mention some important details:

(i)	First, we note that the critical electron density $n_c$ is not universal and changes upon changing the random potential (see Figs.\ref{Figsc} and \ref{Fig11sc}(a) and also Ref.\cite{sar1, sar5}).

(ii)	If we assume that the transition discovered in Si-MOSFETs is a quantum MIT stabilized by electron-electron interactions, then there must be another MIT occurring as the electron density increases and the interaction weakens. But no traces of such behavior were observed (see, \textit{e.g.}, \cite{br1}). However, the possible absence of the second transition was theoretically predicted in \cite{fink1}.

(iii)	Despite a certain success, the renormalization group theory cannot offer any predictions about the structure of an insulator.

\par\bigskip 

\textbf{2.2	Influence of a magnetic field on the metal-insulator transition}\\

In a two-dimensional electron system, magnetic field parallel to the interface acts only on the electrons' spins  and can completely spin-polarize them \cite{vitk}. The spin-polarized electron system in the vicinity of the transition changes its behavior from that typical of a metal to that typical of an insulator and does not exhibit any properties similar to those shown in Fig.\ref{Figsc}. We can see from Fig.\ref{Fig11sc}(b) that the resistance increases with decreasing temperature at all electron densities, although a number of features (the disappearance of non-linearity and the vanishing of the activation energy) demonstrate the transition from an insulator to a metal at the critical density  $n_c=1.155\times10^{11}$ cm$^{-2}$.  Strictly  speaking,  a non-horizontal separatrix separating the metal from insulator does not necessarily mean the absence of a quantum phase transition (this question is discussed in detail in \cite{dolgop,popovic}). Therefore, the determination of the critical density by the sign of the temperature derivative is at least controversial. Below, we use other  criteria  for  finding  the  critical  density $n_c$, for example, by the vanishing activation energy in the insulating phase.

\begin{figure}
\scalebox{1.0}{\includegraphics[width=\columnwidth]{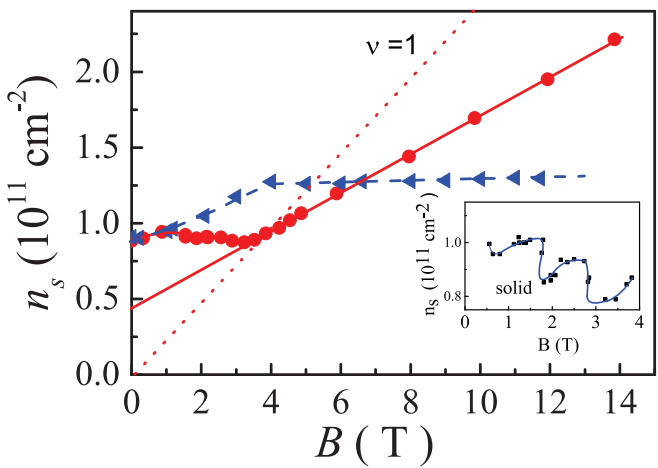}}
\caption{Critical density for the MIT measured for two orientations of a magnetic field: perpendicular to the interface (from \cite{letters}) (dots) and parallel  to  the  interface  (triangles). $n_c (B=0) =0.89\times10^{11}$ cm$^{-2}$. The dashed straight line corresponds to the filling factor  $\nu =1$ . The inset shows critical density oscillations on an expanded scale  \cite{mason}.}
\label{FigBnorm}
\end{figure}

Figure \ref{FigBnorm} shows the position of the MIT on the $(B,n_s)$ plane in a magnetic field normal and parallel to the two-dimensional electron gas. In the parallel orientation of the magnetic field, the transition point is independent of the angle between the current and field, which again confirms that the magnetic field in this case affects only the spins of electrons. (The contribution of orbital effects to the magnetoresistance in Si-MOSFETs is rather weak: in \cite{vol}, a magnetoresistance anisotropy of about 5\% was observed in the parallel field for different current directions with respect to the field.)

In Fig.\ref{FigBnorm}, in the region under the solid line in the perpendicular magnetic field and under the dashed line in the parallel field, the electron system is an insulator, while above these lines it is a metal. Two points must be noted. Firstly, the critical density in the parallel magnetic field gradually increases by approximately a factor of 1.5 and then ceases to increase with increasing magnetic field. Secondly, the behavior of the critical density proves to be quite different in the normal and parallel magnetic fields.

The increase in the critical density in the parallel magnetic field is related to the spin polarization of electrons. In a strong magnetic field (exceeding 4~T in Fig.\ref{FigBnorm})), electrons are completely spin-polarized, and the critical density is independent of the magnetic field. The transition from a spin-polarized electron insulator to a spin-polarized metal was observed in \cite{popovic2}.

The last statement, as well as the scale of the effect and the resistance behavior in the metallic regime, agree with the concept of the MIT being due to the electron-electron interactions  \cite{sh12}. In the metallic phase, a high-viscosity regime appears, which was called a metal glass in Ref.~\cite{popovic3}. A similar regime, although in a considerably narrower density range, can also exist in zero magnetic field. In any case, it is unmistakably observed in strongly disordered Si-MOSFETs  \cite{popovic}.
 
The linear increase in the critical density in the initial region in Fig.\ref{FigBnorm} in the framework of these concepts should correspond to the behavior of the field corresponding to complete spin polarization, which, even if not quite exactly, corresponds to the experiments. Nevertheless, the correctness of the description of the MIT in most perfect Si-MOSFETs based on calculations in \cite{sh12} is doubtful. These calculations correspond to the Anderson transition rather than to a quantum transition. In addition, the transition point in the calculations is determined by extrapolation from the metallic region.

The linear dependence of the critical density in a strong normal magnetic field can be understood based on considerations presented in \cite{loz}. The normal magnetic field reduces the amplitude of zero vibrations of electrons in an insulator ($\propto B^{-1/2}$). According to the Lindemann's criterion, the critical electron density is determined by comparing the amplitude of zero vibrations and the inter-electron distance  ($\propto n_s^{-1/2}$). The extrapolation of the straight line to zero magnetic field specifies the number of localization centers equal to $4\times10^{10}$ cm$^{-2}$ for the data in Fig.\ref{FigBnorm}.
 
In the initial region, where quantization is insignificant, curves for the normal and parallel magnetic fields coincide. But as the magnetic field is increased further, the critical electron density of the MIT in the normal field does not increase, in contrast to the case of the parallel field, and even somewhat decreases, exhibiting small oscillations (see the inset in Fig.\ref{FigBnorm}).

In strong quantizing magnetic fields, each of the quantum levels has a band of delocalized states. As the magnetic field decreases, the energy of the delocalized states decreases with density following the corresponding filling factor. Therefore, the MIT in a strong field should occur when the filling factor is smaller than unity (see Fig.\ref{FigBnorm}). In a weak magnetic  field  ($\omega_c \tau \simeq 1$), delocalized states detach from their quantum levels and, being topologically protected, merge, without decreasing their energy with a further decrease in the field  \cite{khmel,dolgopolov}. 

The boundary oscillations in this region were explained by the oscillations of the chemical potential  under conditions of the quantum Hall effect  \cite{pudalov1,dolgopolov1}. It was assumed that both phases can coexist at the MIT boundary. In this case, the chemical potentials of the phases must be equal at the transition point. The chemical potential of the insulator changes gradually with energy, whereas the chemical potential of the metal in a quantizing magnetic field oscillates, resulting in boundary oscillations.

\par\bigskip

{\bf 2.3. Electron properties in the depth of the Fermi distribution} \par\bigskip

{\bf 2.3.1. Complete spin polarization field}\\

\begin{figure}
\scalebox{0.8}{\includegraphics[width=\columnwidth]{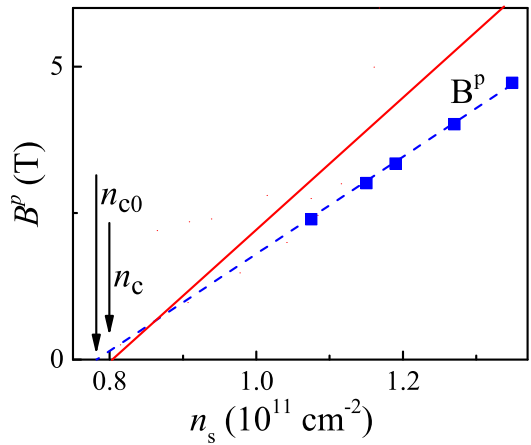}}
\caption{Dots: total spin-polarization field as a function of the electron density (from \cite{Sh}). 
The dashed line is a least-square fit of experimental points. The solid line is the expected result for noninteracting electrons in the model presented in the text.}
 \label{Fig1}
\end{figure}

 {\it Experiment.} The field of the complete spin polarization, $B^p$, parallel to the interface, depends linearly on the electron density \cite{Sh,vitk1,pud} and vanishes when extrapolated to zero at a finite electron density  $n_{c0}$ (Fig.\ref{Fig1}).
 The linear dependence, which was first established from transport measurements, was later confirmed in independent experiments 
 \cite{anis}.
 It was shown in Ref.\cite{pud} that worsening of the sample quality did not change the slope of the linear dependence but increased  $n_{c0}$. Since $n_{c0}$ proved to be rather close to $n_c$ in all experiments, it was assumed that the number of mobile electrons was not equal to $n_s$ but rather to  $n_s-n_c$.

\begin{figure}
\scalebox{0.9}{\includegraphics[width=\columnwidth]{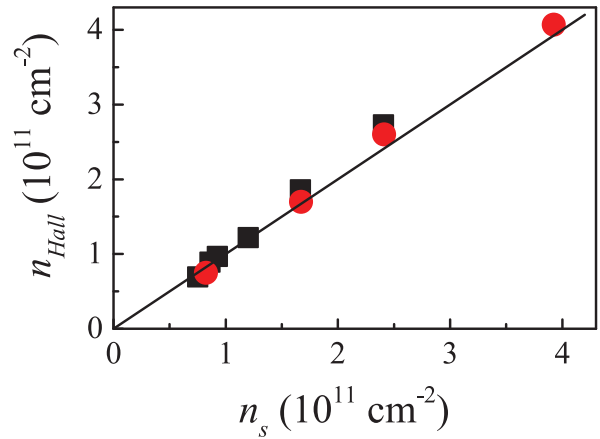}}
\caption{Electron density found from Hall effect measurements in a weak magnetic field as a function of the total electron density determined by quantum oscillations in a strong magnetic field. Dots: $B= 0.3T$, squares: $B=0.1$ T. The solid line corresponds to  $n_{Hall}=n_s$. }
 \label{FigHall}
\end{figure}

To rule out such a possibility, the electron density $n_{Hall}$ was measured in \cite{Sh}  by the Hall effect in a weak magnetic field (Fig.\ref{FigHall}). The electron density measured in these experiments turned out to be coincident within the experimental accuracy with the total electron density $n_s$ determined, as usual, by the Shubnikov-de Haas effect in a strong magnetic field. Below, we will show that such an experiment does not necessarily rule out the existence of a ``tail'' in the density of states containing localized electrons.

The condition for the total spin polarization can be formulated as follows: the Fermi energy, measured from the bottom of the electron subband for electrons gaining the energy $\varepsilon _F ^p$  in the magnetic field, equals to $\mu_B g B^p$ (here $\mu_B$ is the Bohr magneton and  $g$  is the Land\'e $g$-factor taken at the energy  $\varepsilon _F ^p$).  The constant slope of the experimental dependence in Fig.\ref{Fig1} means that $\mu_B g  D_T^p=$~const. Here, the Land\'e $g$-factor and the thermodynamic density of states, $D_T^p$, of the totally spin-polarized  electron  gas  are,  in  principle,  functions  of  the
electron density. Taking into account that the $g$-factor weakly changes with the electron density in the metallic phase,
the constant slope means the absence of the renormalization of the thermodynamic density of states in the totally spin-polarized electron system in (100) Si-MOSFETs.

The discovery of a finite critical density even in the most perfect silicon structures was interpreted as a manifestation of the possible spontaneous spin polarization in a strongly interacting electron gas or at least as magnetic-field-induced instability \cite{Sh,sh1}.

\begin{figure}
\scalebox{0.8}{\includegraphics[width=\columnwidth]{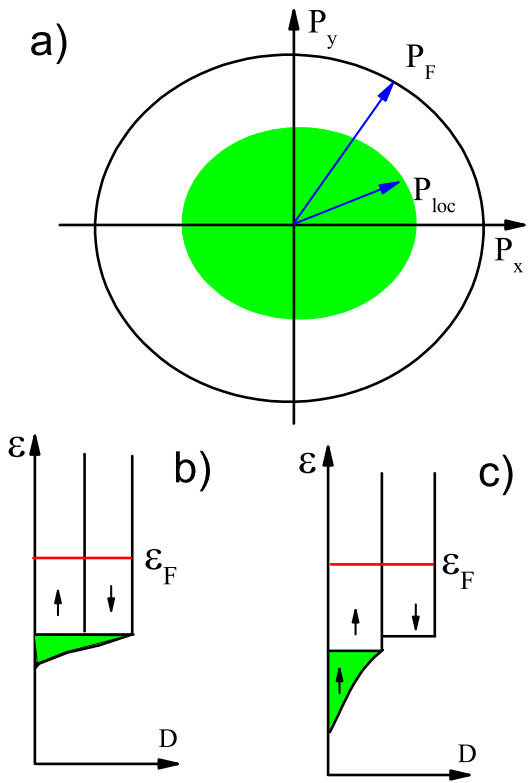}}
\caption{(a) Fermi surface with localized states inside (green filling). (b) Density of states 
 $ D(\varepsilon)=\frac{dn}{d\varepsilon} $ as a function of energy. Localized states (green filling) in the absence of a magnetic field are not spin-polarized. (c) Density of states as a function of energy in a weak magnetic field. Localized states are completely spin polarized.}
\label{Fig2p}
\end{figure}

{\it  The naive model.} The behavior of the total spin polarization field, similar to that shown by dots in Fig.\ref{Fig1}, can also be realized in a two-dimensional system of non-interacting electrons. Indeed, we assume that some of these electrons are localized  (Fig.\ref{Fig2p}).  In zero magnetic field, two spin subbands are filled equally, each of them consisting of two valley subbands. The tail of spin-localized states is not polarized.

 It is important that 

(i) the first delocalized electron has a finite quasimomentum and energy (Fig. \ref{Fig2p}a)
\begin{equation}
p_{loc}= \hbar (\pi n_c)^{1/2},\\   \varepsilon(n_c) = p_{loc}^2/2m^* ,
\label{eqp}
\end{equation} 
 and 

(ii) the number of strongly localized electrons is independent of  $n_s$ ($n_s>n_c$).

Equation  (\ref{eqp}) cannot be proved. It is an assumption and must be verified experimentally. The question about the possibility of verifying this equation is considered in the next sections.

In a weak magnetic field and at a sufficiently low temperature,  all  electrons  in  the  ``tail''  are  spin  polarized (Fig.\ref{Fig2p}(c)). This is possible,  for example, for  single localized spins \cite{gold,gold1}. The analog of Fig.\ref{Fig2p}(a) cannot be drawn for Fig.\ref{Fig2p}(c). 

\begin{figure}
\scalebox{0.9}{\includegraphics[width=\columnwidth]{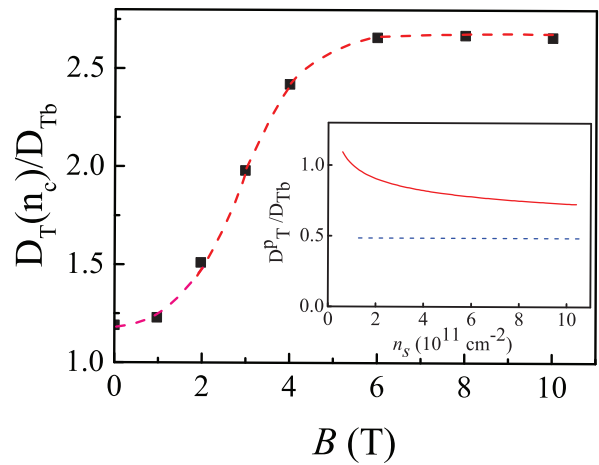}}
\caption{Thermodynamic density of states in the vicinity of the critical density in the insulating phase in a parallel magnetic field (from \cite{sh11}.) The inset is the thermodynamic density of states for completely spin-polarized electrons in the metallic phase for experimental data in Fig.(\ref{Fig1}) (dashed curve) and results of  \cite{waintal} (solid curve).}
\label{FigW1}
\end{figure}

An attempt to measure the thermodynamic density of states in a spin-polarized localized electron system  \cite{sh11} has led to an entirely unexpected result, shown in Fig.\ref{FigW1}: the thermodynamic density of states in a spin-polarized electron system in the insulating phase turned out to be almost three times higher than that for spin-unpolarized electrons and almost six times greater than for spin-polarized electrons.

The polarization of mobile electrons begins at densities $n_s>n_c$. In other words, in the model considered here,  $n_c=n_{c0}$, and the slope of the straight line is determined by the condition
\begin{equation}
\frac{dB^p}{dn_s}=\frac {2}{ \mu_B g  D_{Tb}} = \frac{\pi \hbar^2}{m^*\mu_B g} ,
\label{eqD}
\end{equation}
where  $D_{Tb}=2D_{Tb}^p =\frac{dn_s}{d \varepsilon_F}$ is the thermodynamic density of states of the spin-unpolarized electron gas with band parameters. The corresponding dependence is shown by the solid straight line in Fig. \ref{Fig1} for  $m^*=m_b$ ($m_b$ is the band  electron mass), $g=2$. We can see from Fig.\ref{FigBnorm} that the values of  $n_c$ and $n_{c0}$ virtually coincide but the slope of the experimental straight line is smaller than expected. The slopes can be matched assuming that the $g$-factor exceeds the band $g$-factor: $g=1.3 g_0 =1.3\times2$. Indeed, the measured values of the  Land\'e $g$-factor always exceed  $1.5-1.7\times g_0$  \cite{gersh}, $1.5\times g_0$ \cite{kr}, $1 - 1.3 \times g_0$ \cite{kr1}. All these values were obtained from measurements in a normal magnetic field and therefore it is not obvious that the same values of the Land\'e $g$-factor are also applicable to the magnetic field parallel to the interface.

We note that the presence of the localized states satisfying Eq.(\ref{eqp}) cannot be established in transport measurements in the metallic phase, where all the properties are determined by the close vicinity of the Fermi surface. In particular, measurements of the Hall resistance in weak magnetic fields yield the total electron density.

The assumption that the electron system undergoes a transition to the totally spin-polarized state is also well founded because, for example, a system of localized magnetic moments is already polarized at zero temperature in an infinitely weak magnetic field.

With minor changes, these considerations can be applied to a system of strongly interacting electrons, assuming that we are dealing with single-particle electron states. The average mass of quasiparticles in this model is independent of (or weakly depends on) the electron density. Probably, such a model was used to interpret experimental results in Refs.\cite{pud,kn}.

{\it The alternative naive model} consists of the following. It is assumed that when the critical density $n_c$ is exceeded, all electrons become mobile and the energy of a totally polarized electron gas vanishes at $n_c$:
\begin{equation}
p(n_s)= \hbar (\pi n_s)^{1/2},   \varepsilon_F^p(n_s) = p^2/2m^* \propto (n_s-n_c)
\label{eqp1}
\end{equation} 
\begin{figure}
\scalebox{0.9}{\includegraphics[width=\columnwidth]{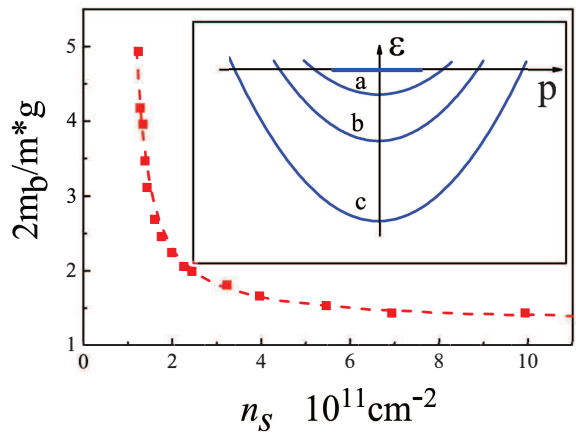}}
\caption{Electron density dependence of the average single-particle mass $m^*$, multiplied by the Land\'e $g$-factor at the Fermi level for spin-polarized electrons (from \cite{Sh}.) The inset shows single-particle electron spectra for three electron densities: (a) $4n_s$, (b) $2n_s$, (c) $n_s$. Energies are counted from the Fermi level.}
\label{Figchi}
\end{figure}
The assumed single-particle spectrum of a completely spin-polarized electron gas is shown schematically in the inset to Fig.\ref{Figchi}. The spectrum is quadratic with the average effective mass  $m^*$. The constant value of the thermodynamic density of states and the vanishing of the total spin polarization field at $n_c$ cause the divergence of   $m^*$, as shown in Fig. \ref{Figchi}.

This model was  used earlier in a number of papers beginning with  \cite{Sh} and ending with recent study  \cite{sh1}.

Both models have a number of disadvantages. First, quasiparticles are treated as free particles even in the depth of the Fermi distribution. Neither model can explain the shift of the critical density as the total spin polarization sets in. Finally, they do not consider the physical properties of an insulator.

\par\bigskip

{\bf 2.3.2. Thermodynamic density of states}\\

As mentioned above, the thermodynamic density of states is one of the parameters that can be inferred from experiments. Experimental data for a completely polarized system of mobile electrons can be compared with the results of numerical Monte Carlo simulations  \cite{waintal}. The solid curve in Fig.\ref{FigW1} shows the thermodynamic density of states of an ideal two-dimensional Si-MOSFET electron system with the electron scattering ignored, calculated using the results in  \cite{waintal}. The dashed line in the inset shows the thermodynamic density of states obtained from experimental data in Fig.\ref{Fig1} using the value $g=1.3\times g_0$. We can see from Fig.\ref{FigW1} that the calculated thermodynamic density of states weakly depends on the electron  density  for $n_s>3\times10^{11}$ cm$^{-2}$,   asymptotically approaching the dashed line  as  $n_s \rightarrow \infty$. The complete spin polarization field calculated in the same density region is consistent with experiment. According to  \cite{waintal}, the consideration of a finite mean free path of electrons reduces the growth of the thermodynamic density of states at minimal electron densities, such that the thermodynamic density of states approaches a constant in the density range of interest to us.  

Comparison of the data in Fig.\ref{FigW1} with that in the inset shows that the thermodynamic density of states in the parallel magnetic field experiences a jump at the density  $n_c$, increasing under conditions of the complete spin polarization
in the insulating phase. 

Information on the thermodynamic density  of  states  can be obtained from the capacitance measurements. Indeed, by measuring capacitances in the absence of a magnetic field and in a field parallel to the plane of a two-dimensional system, we determine their difference  $\Delta C (n_s)$ in the region of complete spin polarization,:
\begin{equation}
\frac{\Delta C}{C}= C_0(Ae^2)^{-1}(\frac{1}{D_{Tpol}}-\frac{1}{D_T}).
\label{eqC}
\end{equation}
where $A$ is the gate area. For an electron density lower than $1.2\times10^{11}$ cm$^{-2}$, the  electron  system  in  a  magnetic  field becomes an insulator, resulting in an increase in  $|\Delta C|$. At the electron density $2.25\times10^{11}$ cm$^{-2}$, 
the transition to the completely spin-polarized state from magnetoresistance measurements is detected.

\begin{figure}
\scalebox{0.9}{\includegraphics[width=\columnwidth]{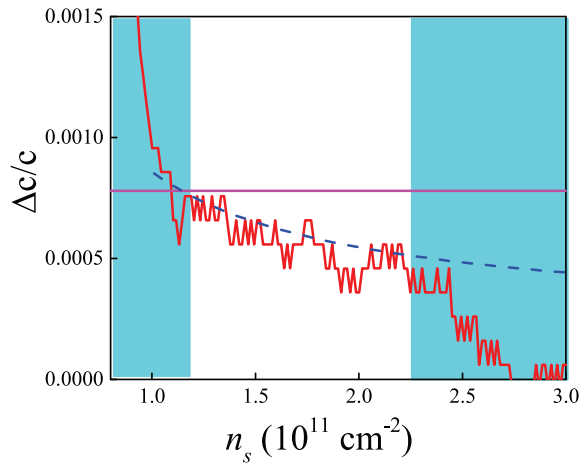}}
\caption{Difference of capacitance measured in the absence of a magnetic field and in  9.9~T parallel magnetic field. The solid straight line shows the level expected for the electron system without interaction. Experimental data should be considered only in the unshaded region. The dashed curve is a fit of the capacitance with Eq.(\ref{eqC}) taking the possible change in the thermodynamic  density  of  states   $D_T \propto \frac{n_s}{n_s-n_c}$ \cite{golddol}. The  last equation should be regarded as a purely empirical one, properly describing experimental data.  }
\label{FigC}
\end{figure}

We can see from Fig.\ref{FigC} that the measured capacitance difference decreases as the electron density increases. This is possible only if the thermodynamic density of states behaves  as  
 $D_T \propto \frac{n_s}{n_s-n_c}$ and $D_{Tpol} = $const. \footnote{The equation for $D_{Tpol}$, obtained in  \cite{D}, is based on a misunderstanding: chemical potentials entering Maxwell relation and screening were measured relative to different levels}. Such  a behavior was reported in  \cite{golddol} based on the analysis of data on the dependence of the elastic relaxation time on the electron density. The fitting of experimental data with this equation and Eq.(\ref{eqC}) is shown by the dashed line in Fig.\ref{FigC} for $D_T =D_{Tb} \frac{n_s}{n_s-n_c}$ and
 $D_{Tpol}= D_{Tb}/1.3$. The  result obtained should be verified by measurements in stronger  magnetic fields.
 
\par\bigskip

{\bf 2.3.3.  Electron magnetization in the metallic phase}\\

Studies of the electron magnetic moment were initiated in \cite{rez}. Because the direct measurement of the magnetic moment of a two-dimensional electron system is difficult due to its smallness, the quantity  $\frac{\partial \mu}{\partial B}$, equal to $-\frac{\partial M}{ \partial n_s}$ according to Maxwell relation, was measured in experiments.  (Here $\mu$ is the chemical potential of the electron system and $M$ is the magnetization.) To   obtain  the dependence   $M(n_s)$, the  measured  quantity  should be integrated over the electron density. However, to do this, it is necessary to use some point with the known magnetic moment as the initial point or to measure $\frac{\partial \mu}{ \partial B}$ in the insulating region at low electron densities for the most perfect samples, which is not easy at low temperatures. Below, we present results obtained by the first \cite{anis} and the second methods \cite{rez,siv,ten,kunts}.
\begin{figure}
\scalebox{1.0}{\includegraphics[width=\columnwidth]{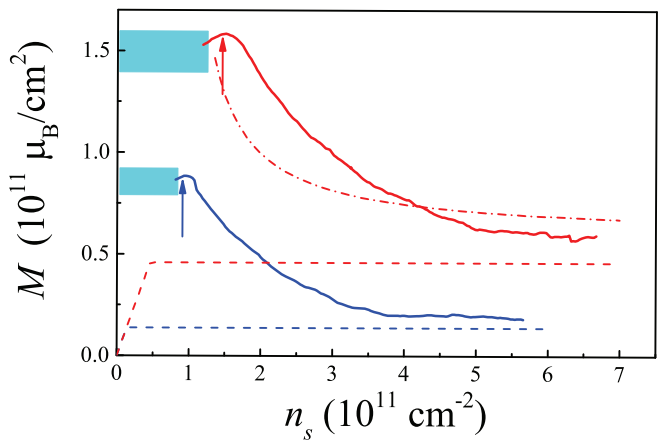}}
\caption{Electron density dependences of the magnetic moment of
the unit area in the 5~T (top solid curve) and 1.5~T (bottom solid curve) parallel magnetic fields, $T=0.4$ K (from \cite{anis}). The dashed line shows the expected behavior of the magnetic moment in the same fields for a free-electron gas with band parameters. The arrows show densities corresponding to the total spin polarization (from data in Fig.\ref{Figchi}). The dashed-dotted curve is calculated in the parabolic dispersion approximation with the same average effective mass of electrons with different spin orientations. The Land\'e $g$-factor and mass correspond to Fig.\ref{Figchi}. Insulating regions are colored.
}
\label{FigM}
\end{figure} 
 
The magnetic moment as a function of the electron density for two magnetic fields is shown in Fig.\ref{FigM}. The curves shown in this figure were obtained from original curves by integrating, assuming that interaction can be disregarded at the maximal electron density and the band mass and the Land\'e factor  $g=1.3g_0$ can be used for calculating the magnetic moment. This procedure is justified by the fact that near the maximum, the magnetic moment expressed in Bohr magnetons coincides with good accuracy with the total number of electrons (the discrepancy does not exceed  $2\times10^{10}$ cm$^{-2}$).

For comparison, the dashed lines in the same figure show the  dependences of the magnetic moment expected for a gas of free electrons with band parameters.  We can see that interaction plays a considerable role and significantly modifies the dependence $M(n_s)$. In addition, the dashed-dotted curve shows the expected dependence for a gas of interacting electrons, assuming that electrons with oppositely oriented spins have a parabolic spectrum with a mass  $\propto \frac{n_s}{n_s-n_c}$. We can see that the spectrum of a partially polarized electron system is not quadratic, and a comparison of the values of two solid curves at an arbitrary fixed density suggests that the degree of spin polarization is approximately proportional to the magnetic field.

Recently, a method for measuring the magnetization in an insulator was proposed in Ref.~\cite{kunts}. Measurements are possible at a low but finite conductivity at a finite temperature. The method is based on the fact that MOSFET charging occurs in the same way under modulation of the gate voltage and parallel magnetic field. In the first case, some effective capacitance is measured (which depends on the conductivity and does not exceed the sample capacitance), and in the second case the value of $\frac{\partial \mu}{\partial B}$, related to the same area as that the measured capacitance is determined. The only unwelcome
feature of the method is that the magnetic moment measured is related to the near-contact region of the sample.

\begin{figure}
\scalebox{1.0}{\includegraphics[width=\columnwidth]{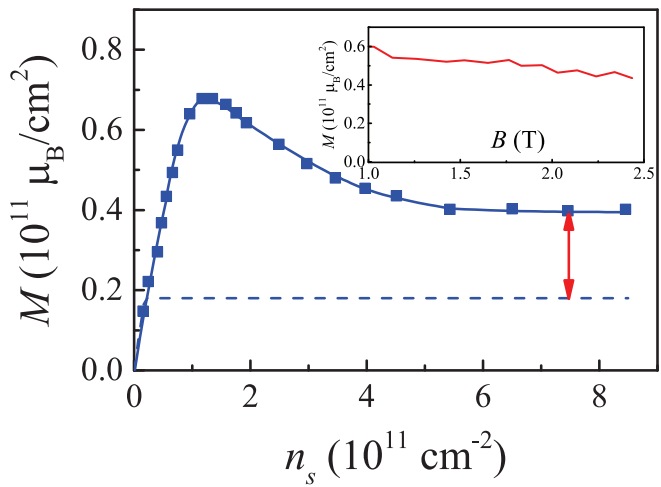}}
\caption{Density dependence of the magnetic moment of the unit area in the 2~T parallel magnetic field at the temperature 1.7~K (squares). The dashed straight line shows the expected behavior of the magnetic moment in the same fields for a free-electron gas with band parameters. The arrow corresponds to the maximum number of localized electrons in the metallic phase. The inset demonstrates the dependence of the magnetic moment on the magnetic field at the electron density  $4.3\times10^{11}$ cm$^{-2}$.  Adapted from \cite{ten}. }
\label{Figpud1}
\end{figure}

One of the results in  \cite{ten}, obtained by integrating the magnetic moment starting from the zero electron density, at which the magnetic moment of the electron system is zero, is shown in Fig.\ref{Figpud1}. The curve presented in the figure is consistent  in  the  overlap  region   ($n_s<6\times10^{11}$ cm$^{-2}$) with data obtained earlier by a different method \cite{rez}, for which the magnetic moment is approximately proportional to the magnetic field and is therefore mainly caused by mobile electrons. As the electron density is increased further, the magnetic moment ceases to decrease (Fig.\ref{Figpud1}).  The level at which the dependence $M(n_s)$ is saturated depends weakly on the temperature and the magnetic field strength (see the inset in Fig.\ref{Figpud1}), which is without doubt related to the presence of localized electrons in the metallic phase and which allows estimating their number. Indeed, because the magnetic moment at the saturation depends on the temperature very weakly, the maximum number of localized electrons is equal to the difference between the saturation level and the expected level  for  a  free  electron  gas.  It  is   $2\times10^{10}$ cm$^{-2}$,  whereas $n_c \approx 8.5\times10^{10}$ cm$^{-2}$  for   the   sample   under   study.   This number is not universal. It changes from sample to sample and can also change from cooling to cooling, even for the same sample.

\begin{figure}
\scalebox{1.0}{\includegraphics[width=\columnwidth]{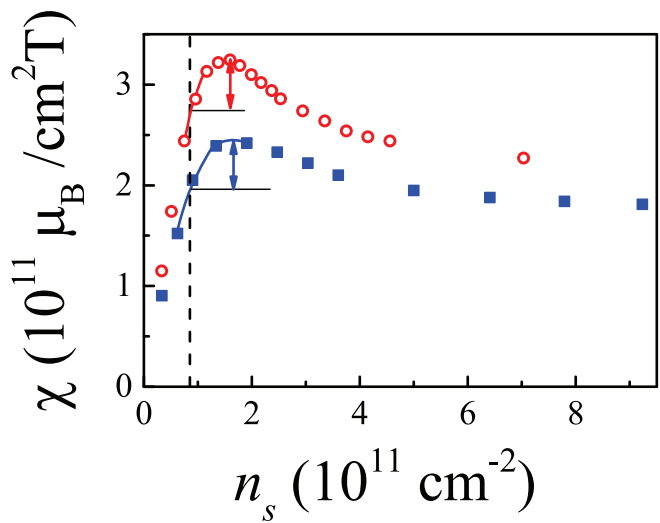}}
\caption{Dependences of the magnetic susceptibility in the zero magnetic field on the electron density for two samples from different manufacturers (dots: samples made in Russia, squares: a sample made in the Netherlands). The vertical dashed straight line indicates the MIT. $T=1.7$~K. From \cite{ten}. }
\label{Figpud2}
\end{figure} 

Much effort has been made to study the magnetic susceptibility, \textit{i.e.}, the quantity  $\chi=\frac{\partial M}{\partial B}(B=0)$ (see Fig.\ref{Figpud2}). A comparison of the result presented in Fig.\ref{Figpud2} with a rough estimate of the expected magnetic susceptibility according to the data in  Figs.\ref{FigM} and \ref{Figpud1}  shows that the measured susceptibility exceeds the expected one by almost an order of magnitude. This is possible only if the magnetic susceptibility is caused by the initial stage of the rearrangement of the ``tail'' of the density of states (see Fig. \ref{Fig2p}) and is not directly related to the properties of delocalized electrons.

We note two features of the curves in Fig.\ref{Figpud2}. First, the curve obtained for a sample from the Netherlands lies below the corresponding curve for a sample made in Russia. Assuming that with all other parameters being equal, the susceptibility in the metallic phase is proportional to the number of localized electrons, we can conclude that the number of localized electrons in the Russian sample is greater by 25\%.

Second, the susceptibility in a small density interval continues to increase with increasing $n_s$ in the metallic phase, and this increase is virtually the same for both samples (arrows in Fig.\ref{Figpud2}).  

\par\bigskip

{\bf 2.4. Electron properties at the Fermi level}

\par\bigskip

{\bf 2.4.1	Temperature dependence of the conductivity}\\

In the absence of a magnetic field, the conductivity of a two- dimensional electron system linearly depends on temperature in some temperature range. Such a behavior of the conductivity was predicted by two different models  \cite{dolg,al,stern} and experimentally demonstrated in  \cite{dor}. The temperature interval in which a linear dependence is expected is determined by the condition:
\begin{equation}
\hbar/\tau \ll kT \ll p_F v_F ,
\label{eqT1}
\end{equation}
where  $k$ is the Boltzmann constant and  $p_F$ and $v_F$ are the electron momentum and velocity on the Fermi  surface.  The
left inequality in relation (\ref{eqT1}) corresponds to the ballistic regime \cite{al}. It appears in the alternative model  \cite{dolg}, as the restriction on energy in the regime where the screening parameter is washed out by collisions.

It is important for us that regardless of the model, the conductivity in the linear region is determined by the relation
\begin{equation}
\frac{\sigma (T)}{\sigma(0)}= 1- AkT,
\label{eqT2}
\end{equation}
where $A \propto (p_Fv_F)^{-1} \propto m^*_F/n_s$. Here, we introduce, in a standard way, the single-particle mass on the Fermi surface as  $m^*_F = p_F/v_F$.

\begin{figure}
\scalebox{0.9}{\includegraphics[width=\columnwidth]{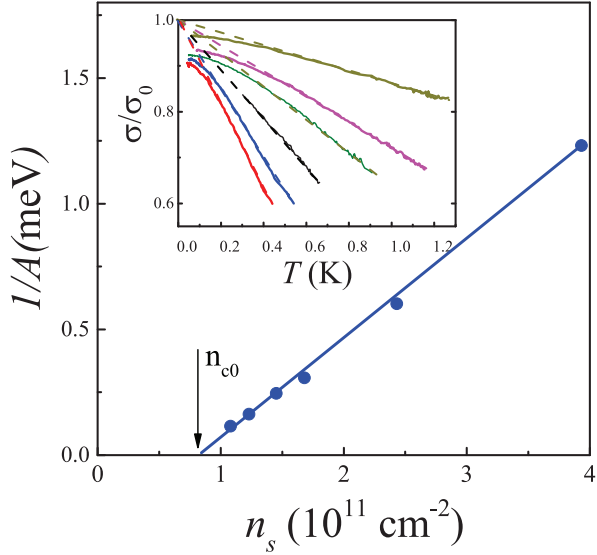}}
\caption{Inverse  slope  $1/A$  of  the  temperature  dependence  of   the normalized conductivity as a function of the electron density. The inset  shows the temperature dependence of the normalized conductivity at the electron densities (from top down)  $2.4, 1.68, 1.45, 1.23, 1.08, 1.01 \times 10^{11}$ cm$^{-2}$. From \cite{kr}. }
\label{Figtemp}
\end{figure}

Examples of the temperature dependence of the conductivity on the metallic side of the MIT are shown in the inset in Fig.\ref{Figtemp}. For each of the electron densities, the temperature dependence of the conductivity has a linear region that allows determining  $A(n_s)$. This dependence is shown in Fig.\ref{Figtemp}. The dependence is linear with  good accuracy and can be extrapolated to a finite density coinciding for the sample under study with $n_{c0}$ and $n_c$. The  linear  dependence  means  that  $m^*_F \propto \frac{ns}{n_s-n_c}$, similarly to the behavior of the average mass of spin-polarized electrons, and by extrapolation diverges at the same density where the Fermi energy of spin-polarized electrons reaches the bottom of the electron subband (see Fig.\ref{Fig1}).

\par\bigskip

{\bf 2.4.2. Thermopower}\\

\begin{figure}
\scalebox{0.9}{\includegraphics[width=\columnwidth]{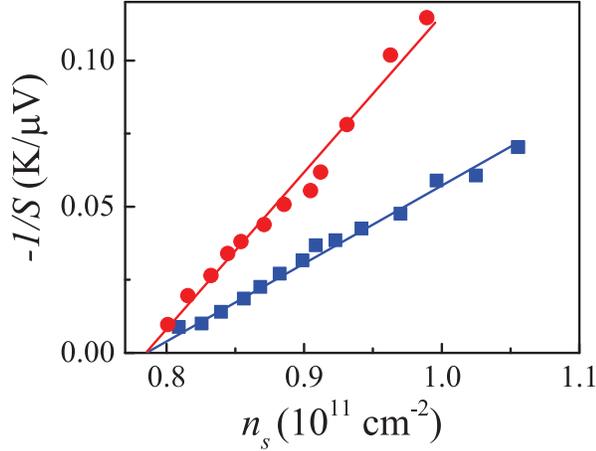}}
\caption{Dependences of the inverse thermopower on the electron density at 300~mK (dots) and 600~mK (squares). The straight lines drawn through the points are extrapolated to the density $n_{c0}$. From \cite{kr}.}
\label{Figtherm}
\end{figure} 

The alternative method for studying the electron properties in the vicinity of the Fermi level in Si- MOSFET involved measurements of the thermopower $S= -\Delta V/\Delta T$ \cite{Bo}, where $\Delta V$ is  the  potential  difference caused by the temperature difference  $\Delta T$ for a constant thermal flow directed along the electron layer. The creation of such a thermal flow and control of its constancy are the major experimental challenges in measuring the thermopower at low temperatures.

In the case of non-interacting electrons (taking valley degeneracy into account), the thermopower is equal to 
\begin{equation}
S= - \frac{2 \pi k^2 m_b T}{3e \hbar^2 n_s}.
\label{eqterm}
\end{equation}
At a low electron density, the elastic relaxation time itself becomes temperature dependent  \cite{dolg,al}, resulting in a correction to Eq.(\ref{eqterm}). On the right-hand side, an additional factor appears depending on the disorder  \cite{flet,Fani,Gos} and
interaction  \cite{gold12}. In addition, for interacting electrons, $m_b$ in Eq.(\ref{eqterm}) should be replaced by   $m^*_F$.
It is expected that the quantity  $1/S$ will be inversely proportional to the temperature and in the simplest case proportional to  $n_s/m^*_F$.
  
Indeed, experiments demonstrate the correct scaling of the thermopower with the temperature and a linear dependence of the inverse thermopower on the electron  density (Fig.\ref{Figtherm}). This implies a constant value of the additional factor in Eq.(\ref{eqterm}) caused most likely by the narrowness of the electron density interval in which the measurements were performed.

The thermopower measurements confirm the dependence  $m^*_F \propto \frac{ns}{n_s-n_c}$ at  minimal  achievable  electron  densities and extend it much closer to the critical density.

\par\bigskip

{\bf 2.4.3. Entropy measurements}\\

Additional information on the properties of the electron system in Si-MOSFET was obtained from entropy measurements  \cite{kunzevich}. Rather complicated experiments involved the study of the response of the chemical potential of the electron system to the temperature modulation $\frac{\Delta \mu}{\Delta T}$, equal to the change in entropy with the opposite sign after the addition of one electron. The entropy $S$ of the unit area of a degenerate non-interacting electron gas ($kT \ll \varepsilon_F$, where $\varepsilon_F$ is the Fermi energy measured from the bottom of the electron subband) is
\begin{figure}
\scalebox{0.9}{\includegraphics[width=\columnwidth]{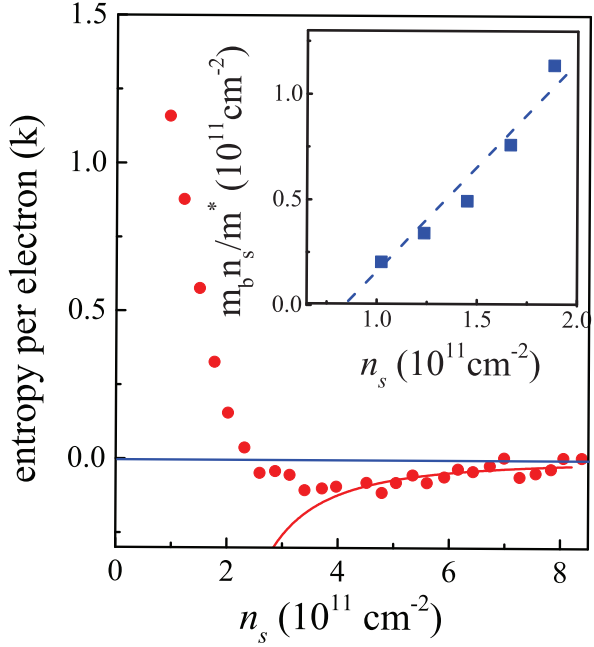}}
\caption{Dependence of  $\frac{\partial \bf S}{\partial n_s}$ on the electron density (in units of the Boltzmann constant $k$) at the temperature 3 K (dots). The solid curve was calculated  by  Eqn. (\ref{eqentr1}) with  the  mass $m^*_F = m_b \frac{n_s}{n_s-n_c}$, where   $n_c=8\times10^{10}$ cm$^{-2}$. The inset shows the result of the analysis of the data shown in the upper half-plane with Eq.(\ref{eqentr2}). The dashed straight line corresponds to the dependence $m^* = m_b \frac{n_s}{n_s-n_c}$.
From Ref.\cite{kunzevich}. }
\label{Figentr}
\end{figure}

\begin{equation}
{\bf S}=k \pi Tg_s g_v m_b/6 \hbar^2
\label{eqentr}
\end{equation}
and is independent of the number of electrons. Therefore, the zero response is expected for non-interacting electrons.
The properties of a degenerate electron gas of interacting electrons are determined by the nearest vicinity of the Fermi level
\cite{pud14}:

\begin{equation}
\frac{\partial \bf S}{\partial n_s}=  \frac{\partial m^*_F}{\partial n_s} k \pi Tg_s g_v/6 \hbar^2.
\label{eqentr1}
\end{equation}
Because the electron mass at the Fermi level increases with decreasing electron density, the negative values of 
$\frac{\partial \bf S}{\partial n_s}$ are expected in the region of the degenerate gas.

The corresponding experimental data are presented in Fig.\ref{Figentr}. Here, the solid curve shows calculations with the electron mass at the Fermi level found previously. At an electron  density  above   $4\times10^{11}$ cm$^{-2}$,  the  calculation  is consistent with experiments. At lower densities, experimental points deviate from the calculated curve and even move to the upper half-plane, because the electron system ceases to be degenerate.

In the opposite case  $kT\geq \varepsilon_F$, the authors of Ref.\cite{kunzevich} analyzed their experimental results using the equation
\begin{equation}
\frac{\partial \bf S}{\partial n_s}= k(\frac{\varepsilon_F /kT}{e^{\varepsilon_F /kT}-1}- ln(1-e^{-\varepsilon_F /kT}))
\label{eqentr2}
\end{equation}
for an ideal gas with a renormalized average  effective  mass $m^*$.

Eq.(\ref{eqentr2}) is valid only in a bounded region of electron densities depending on temperature. For example, in Fig.\ref{Figentr} these are densities above $10^{11}$ cm$^{-2}$ (to avoid the insulator) and below $2\times10^{11}$ cm$^{-2}$ (to remain in the non-degenerate regime). The result of the data analysis is presented in the inset to Fig.\ref{Figentr}. We can see that experimental points in chosen coordinates are close to a straight line with a slope of 45$^0$, extrapolating to a finite electron density.

We note in concluding this section that results obtained in entropy measurements cannot confirm the results in Sections 2.3.3, 2.4.1, and 2.4.2, but do not contradict them either.

\par\bigskip

{\bf 2.4.4.  Shubnikov-de Haas effect }\\

The parameters of an electron system at the Fermi level can be determined from quantum oscillations of resistance (the Shubnikov-de Haas effect) \cite{lif}. Corresponding measurements have been performed by different experimental groups  \cite{butch,rah,klim,pud14} with samples from various manufacturers. The effective mass was found from the Lifshitz-Kosevich relations  \cite{lif}, which give the dependence of the relative magnitude $U$ of quantum oscillations on the temperature and the magnetic field:

\begin{figure}
\scalebox{1.0}{\includegraphics[width=\columnwidth]{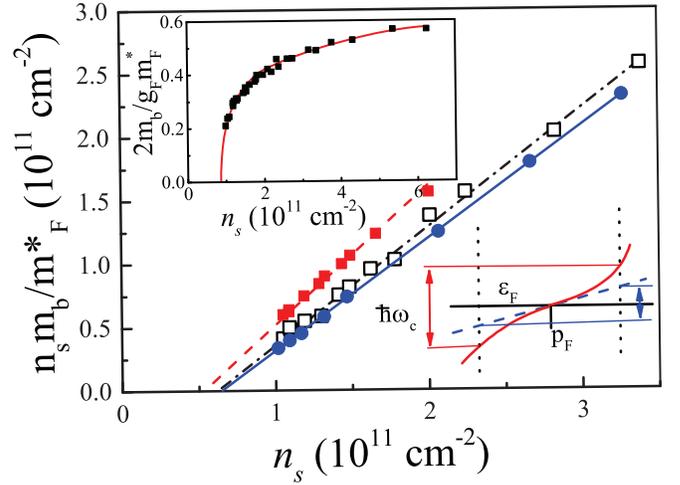}}
\caption{Effective electron mass at the Fermi level extracted from experimental quantum oscillations under the assumption that the Dingle temperature is temperature-independent: unfilled squares  \cite{butch}, dots \cite{rah}, filled squares (the quantum relaxation time is replaced by the transport time)  \cite{butch}. Left inset: the product  $g_Fm^*_F$, measured by quantum oscillation beats  \cite{butch}. The right inset illustrates the decrease in the measured effective mass caused by the nonlinearity of the electron spectrum near the Fermi level. The dashed line is the result of the semiclassical Bohr-Sommerfeld quantization of electron orbits in the $p$-space. The right arrow is the expected cyclotron energy in the case of a linear spectrum. The left arrow is the measured cyclotron energy. The flattening of the single-particle spectrum is connected to the Fermi level and shifts together with it with changing the electron density or spin polarization. The region of a strong spectral nonlinearity also shifts with it. However, the existence of this effect in silicon MOSFETs has not been conclusively proved in the literature so far. }
\label{Figgersh}
\end{figure} 
 
\begin{eqnarray}
U&=&\sum_iU^{LK}_i\cos\left[\pi i\left(\frac{\hbar c\pi n_s}{eB_\perp}-1\right)\right]Z^s_iZ^v_i\nonumber\\
U^{LK}_i&=&4\exp\left(-\frac{2\pi^2ik_BT_D}{\hbar\omega_c}\right)\frac{2\pi^2ik_BT/\hbar\omega_c}{\sinh\left(2\pi^2ik_BT/\hbar\omega_c\right)}\nonumber\\
Z^s_i&=&\cos\left(\pi i\frac{\Delta_Z}{\hbar\omega_c}\right)=\cos\left(\pi i\frac{gm^*_F}{2m_e}\right)\nonumber\\
Z^v_i&=&\cos\left(\pi i\frac{\Delta_v}{\hbar\omega_c}\right),\label{A}
\end{eqnarray}
where $T_D$ is the Dingle temperature, $m_e$ is the free electron mass,  $\hbar\omega_c$ is  the  cyclotron  frequency, $\Delta_Z$ is  the  Zeeman splitting, and $\Delta_v$ is the valley splitting. 

In weak magnetic fields  ($U\ll 1$), the amplitude is determined by the factor $U_1^{LK}$, and two fitting parameters $m^*_F$ and $T_D$ remain in the temperature dependence in Eq.(\ref{A}). The results of such an analysis of the data of \cite{butch} amd \cite{rah} is
shown in Fig.\ref{Figgersh}.

We can see from this figure that data obtained for samples from different manufacturers agree with each other within the experimental accuracy, are described well by a linear dependence in corresponding coordinates, and are extrapolated to the density of $0.66\times10^{11}$ cm$^{-2}$.
 
It was shown in Section 2.4.1 that the transport elastic relaxation time depends on temperature. The questions arise as to whether the temperature dependence can also be manifested in the quantum relaxation time determining the Dingle temperature, and if so, how the effective mass changes after the data analysis with the Dingle temperature depending on $T$.  This effect was roughly estimated in \cite{butch} by replacing the quantum relaxation time in the Lifshitz-Kosevich formula by the temperature-dependent transport time. The result is shown by filled squares in Fig.\ref{Figgersh}. Despite quantitative differences with the previous processing, experimental points again lie on a straight line extrapolated to the density of $0.55\times10^{11}$ cm$^{-2}$. 
 
Replacing the quantum relaxation time with the transport time changes the temperature dependence of the Dingle temperature. The critical electron density found from quantum  oscillations   arguably  lies   in  the  interval   from
$5.5\times10^{10}$ to $6.5\times10^{10}$ cm$^{-2}$. This value is noticeably lower than that found in Sections 2.4.1 and 2.4.2. The reason for the discrepancy  can  be  the  nonlinearity  of  the single-particle electron spectrum near the Fermi level (see the right inset in Fig.\ref{Figgersh}). Indeed, at minimal electron densities, the temperature dependence of oscillations is studied at the third or fourth Landau levels, \textit{i.e.}, under the conditions $\hbar \omega_c \simeq \frac{1}{3}( p_Fv_F/2)$. The nonlinearity of the spectrum for such considerable deviations from the Fermi level can lead to a decrease in the measured effective mass compared with the mass measured directly at the Fermi level.

To avoid misunderstanding, we make an important remark. The amplitude of quantum oscillations is determined exclusively by the vicinity of the Fermi level and is absolutely  insensitive  to  the  difference  between  the band
bottom  and Fermi  energies.  For  this reason,  the measured mass turned out to be insensitive to the spin polarization degree.

The study of the temperature dependence of the amplitude of quantum oscillations in tilted fields  \cite{rah} revealed another important fact about the independence of the electron mass at
the  Fermi  level  from  the  spin  polarization  degree.  This statement was recently confirmed by independent experiments 
 \cite{pud14} and by some raw experimental data  \cite{butch} and
calculations \cite{mas} for a multi-valley electron system in the weak-coupling limit.
 
In a tilted magnetic field, another possibility exists for measuring parameters of the electron system at the Fermi level. By changing the tilt angle (or changing one of the components of the magnetic field with the other component kept fixed), the nodes of quantum oscillations can be observed. It was shown in    \cite{gor} that for a relatively weak electron-electron   interaction   and   zero   temperature,   the
position  of  the  nodes  is  determined  by  the  product   $g_Fm^*_F$. The corresponding experimental data presented in the inset in
Fig. \ref{Figgersh} demonstrate the critical behavior of the product   $g_Fm^*_F$ but cannot be used to accurately determine the critical density because the extrapolation law is unknown.

\par\bigskip

{\bf 2.4.5. Low-frequency resistance noise}\\

Information on electron properties at the Fermi level can be obtained by measuring low-frequency resistance noise. Such measurements were performed with Si-MOSFETs of different qualities \cite{dragana} in a broad temperature range. Below, we consider only the results of low-temperature measurements with the most perfect samples  \cite{dragana1,dragana2}.

It was shown that the low-frequency spectral density of noise, which in the metallic phase is usually proportional to  $1/f$,
changes in the narrow region above  $n_c$ to $1/f^\alpha$ with the exponent $\alpha >1$, increasing with decreasing the electron density. The spectral noise density in this region increases for  $T<3$~K with decreasing temperature. Such a behavior is typical for the amorphous phase (glassy phase).

\begin{figure}
\scalebox{0.9}{\includegraphics[width=\columnwidth]{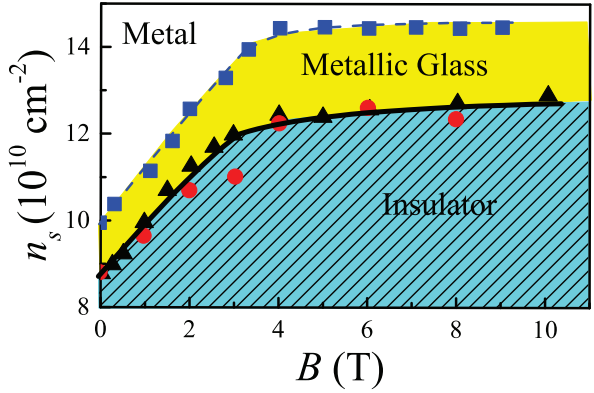}}
\caption{Modified phase diagram for the MIT in a magnetic field parallel to the interface (\textit{cf}. Fig.\ref{FigBnorm}). The hatched region is the insulator, the filled region is the amorphous metal. From \cite{dragana2}.}
\label{Figglass}
\end{figure}

The exponent  $ \alpha$, in a spin-polarized metal, as in the usual metallic phase, is independent of the electron density and is equal to $ \alpha \simeq 0.5$. In the region of transition from a spin-polarized insulator to a spin-polarized metal  \cite{dragana2} $\alpha$ increases, and the density range in which a metallic glass phase exists expands (Fig.\ref{Figglass}).

\begin{figure}
\scalebox{0.9}{\includegraphics[width=\columnwidth]{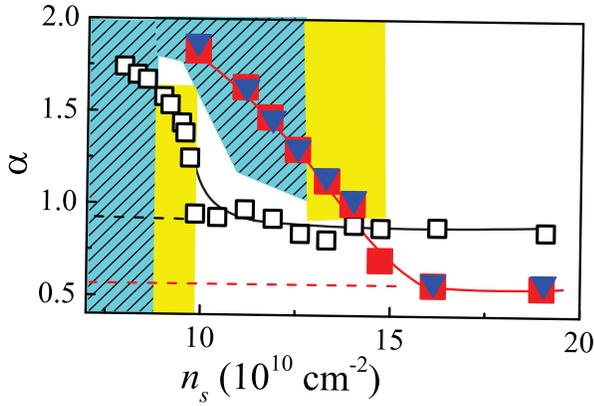}}
\caption{Exponent $ \alpha$ of the frequency dependence of the spectral density of low-frequency noise ($\propto  1/f^\alpha$) 
as a function of the electron density. Unfilled squares: $B=0$, triangles  and red squares: $B=4$~T and $B=9$~T parallel magnetic fields, respectively. The filling corresponds to Fig.\ref{Figglass}. The dashed horizontal straight lines show the saturation level of  $\alpha$ in the metallic phase. From \cite{dragana2}.}
\label{Figalpha}
\end{figure}

While the boundary between an amorphous metal and   an insulator can be determined with good accuracy (see Sections 2.1 and 2.2), the upper boundary (dashed curve) in Fig.\ref{Figglass} is somewhat conventional because of the absence of any criterion for the value of the exponent  $\alpha$ allowing the separation of the metallic phase from the amorphous metal. The accuracy of determining this boundary can be estimated from data presented in Fig.\ref{Figalpha}.

We note that the electron density range where the amorphous metallic phase is observed strongly depends on the sample quality \cite{popovic}, considerably expanding with increasing disorder.

\par\bigskip

{\bf 2.5. Intermediate conclusions}\\

We briefly formulate the results of the experiments presented above.

(i)	The electron mass in the metallic phase at the Fermi level increases as the electron density decreases and is independent of the spin polarization degree. The extrapolation predicts a divergence of mass (the appearance of a flat region in the electron spectrum) at the electron density close to the MIT point in the zero magnetic field for the best of the samples studied.

(ii)	In the metallic phase, a fraction of localized electrons can be retained, their number being dependent on the sample  quality.

(iii)	In the metallic phase, an amorphous metal with long-range correlations of fluctuations can exist in the nearest vicinity of the MIT  \cite{dragana1}. In the most perfect samples, the amorphous metallic region in the zero magnetic field virtually disappears.

( iv ) The energy-averaged electron mass in the metallic phase also increases with decreasing  electron density. More precisely, this statement can be formulated as follows: the distance between the bottom of the electron subband and the Fermi level decreases with decreasing electron density faster than can be expected for a non-interacting electron gas.

(v)	The thermodynamic density of states of spin-non-polarized electrons is proportional to the electron mass at the Fermi level. However, this statement should be additionally verified.

( vi ) The total spin-polarization field is linear in the electron density and is extrapolated for the best samples to zero at the electron density close to the density of the MIT in the zero magnetic field. This means that the thermodynamic density of states of spin-polarized electrons under the condition  $n_s=n_\uparrow$ (where  $n_\uparrow$ is the number of electrons with an energy-advantageous spin orientation) is independent of the electron density.

( vii ) The energy-averaged effective mass of spin-polarized electrons increases  as  the  electron density decreases, demonstrating (by extrapolation) the tendency to diverge at an electron density close to that of the MIT in the zero magnetic field.

\par\bigskip

{\bf 2.6   Electrons in the insulator}

\par\bigskip

{\bf 2.6.1. Low-frequency noise in the insulating phase}\\

In \cite{dragana1,dragana2}, measurements of low-frequency noise were extended to the MIT and even into the insulating phase (see Fig.\ref{Figalpha}). We can see from the figure that neither in the absence of a magnetic field nor in the magnetic field spin-polarizing the electron system were any specific features in the behavior of  $\alpha$ observed at the MIT point. Therefore, the transition occurs between the amorphous metallic phase (with a finite resistance at the zero temperature) and the glassy insulating phase (with the conductivity tending to zero with decreasing temperature).

The amorphous phase of an insulator was considered in the grating model  \cite{dobr1} for spinless electrons. It was shown that in the case of disorder and strong electron-electron coupling, a gapless state appears with a deep lowering of the single-particle density of states, reaching zero at the Fermi level. With such a spectrum realized, variable-range hopping conductivity should be expected \cite{efros}. The temperature dependence of the resistance described by the Efros-Shklovskii law was observed experimentally deep in the insulator  \cite{gesha1,vol2}. In the region closer to the MIT, the usual activation dependence was observed  \cite{sh11}, demonstrating the transition to nearest-neighbor hopping.

The appearance of an intermediate amorphous metallic phase was predicted in Ref.\cite{dobr2}.

\begin{figure}
\scalebox{0.9}{\includegraphics[width=\columnwidth]{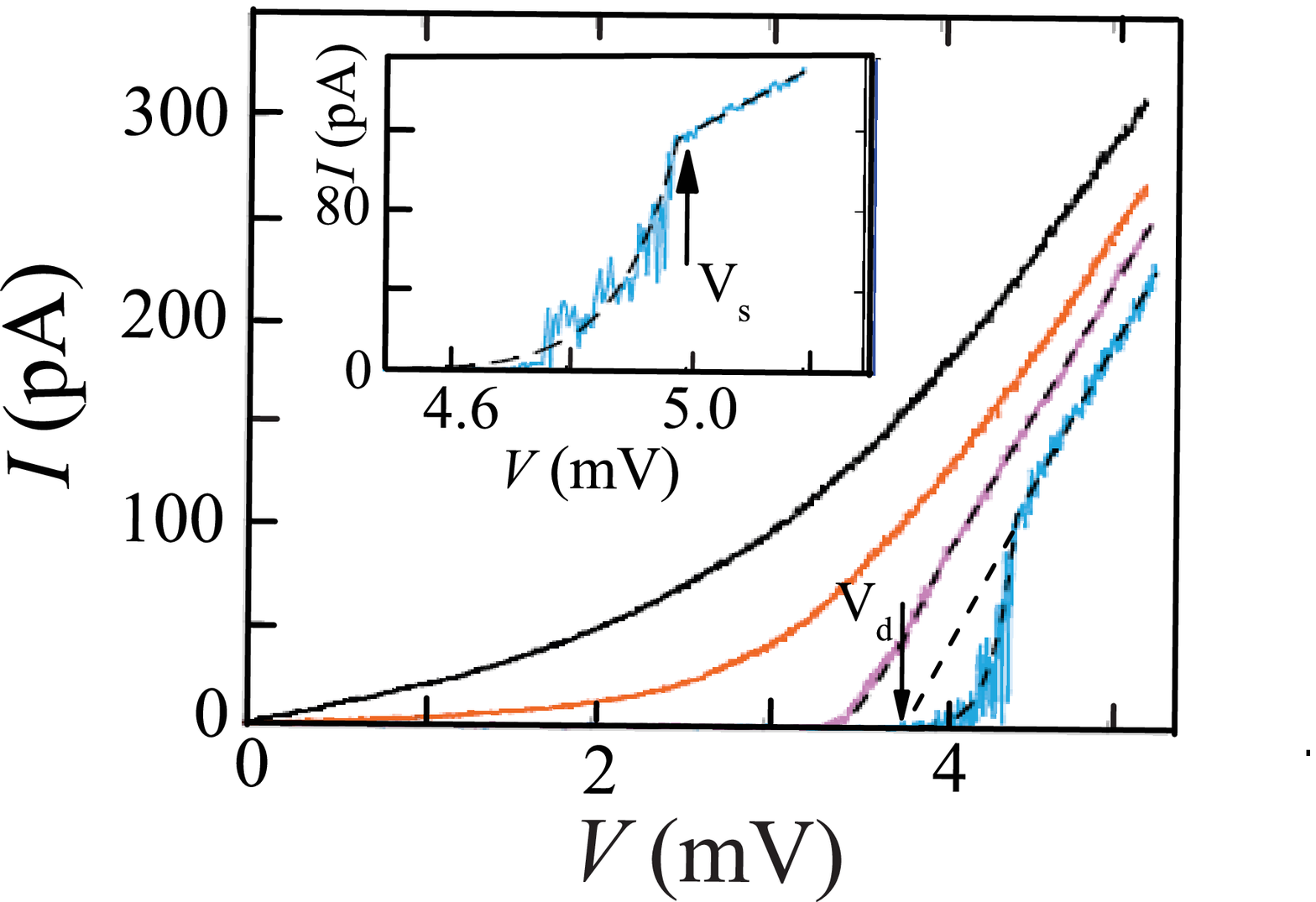}}
\caption{Voltage-current   characteristics   in   the   insulating   phase. $n_s=5.36\times10^{10}$ cm$^{-2}$, temperature  (from  right  to  left):  $60, 140, 200,300$~mK. The inset shows the  $I-V$ at electron density $n_s=5.2\times10^{10}$ cm$^{-2}$  on  the  expanded  scale.  $T =60$  mK.  A  strong increase  in  the  low-frequency  noise  is  observed  in  the  interval  $V_d < V <V_s$. From \cite{miriam}. }
\label{Figsh}
\end{figure} 

Interesting results on low-frequency noise in the insulating phase were obtained in Ref.\cite{miriam}, where nonlinear voltage-current $V-I$ characteristics were studied in the depth of the insulator (Fig.\ref{Figsh}). At a low temperature  ($T \simeq 60 mK $),
 in the linear regime, the current was absent within the experimental accuracy. As the voltage reached a critical value depending on the difference  $n_c-n_s$, the current began to increase dramatically, its increase being accompanied by the low-frequency noise, well observed in the inset in Fig.\ref{Figsh} . Finally, as the voltage reached the second threshold value, the $V-I$ curves became linear and the noise amplitude decreased. Both threshold  voltages decreased with increasing temperature, the noise decreased, and the current appeared in the linear regime at high temperatures. The slope of the linear part of the $V-I$ curves weakly depended on the temperature (see Fig.\ref{Figsh}) and the electron density.

The observed $V-I$s are similar (with the current and voltage axes interchanged) to the well-known $I-V$s characteristics at the depinning of a vortex lattice in type-II superconductors (see, \textit{e.g.}, \cite{lar}). Based on this analogy, we can attempt to describe the experimental curves.

Following the terminology used to describe the properties of the vortex lattice, we introduce two critical voltages: the static voltage  $V_s$ (see the inset in Fig.\ref{Figsh}), corresponding to the onset of the linear dependence of the current on voltage, and the dynamic voltage $V_d$, the result of extrapolating the linear dependence to zero (see Fig.\ref{Figsh}).
 
The region of voltages  $V_d < V < V_s$ is the region  of  collective pinning of an amorphous electron system with a strong inter-particle interaction. In this region, pinning is produced by centers of different strengths, and the electron system can move only due to thermal activation. We note that we are dealing not with the activation of a single electron but with the activation motion of the total electron system or, at least, of a large part of it:
\begin{equation}
I \propto exp [-U(V)/kT].
\label{carr}
\end{equation}
Here $U(V)$ is the activation energy depending on the potential difference applied to the sample.

For voltages exceeding  $V_s$, the electron system moves with friction, which is maximal at spatial points with the greatest pinning force. Therefore,
\begin{equation}
  U_c = eE_s  L,
\label{e}
\end{equation} 
where  $U_c$ is the maximum activation energy in the absence of an electric field,   $E_s$ is the electric field at the voltage $V_S$, and  $L$ is the characteristic distance between the points of maximum pinning. It is the random arrangement of these points that supports the amorphous state of the electron system. The electric current in this region linearly depends on the applied voltage
\begin{equation}
I = \sigma_0 (V-V_d).
\label{I}
\end{equation}
where $\sigma_0$ is a coefficient with the dimension of inverse resistance.

Because the activation energy is described by the equation

\begin{equation}
U(V) = U_c-eEL = U_c (1-V/V_s),
\label{eqU}
\end{equation}
the current at  $V<V_s$ is equal to
\begin{equation}
I = \sigma_0 (V-V_d)  \exp \left[\frac{-U_c(1-V/V_s)}{kT} \right].
\label{V}
\end{equation}

Fitting the experimental curves with Eqs.(\ref{I},\ref{V}) is shown by dashed curves in Fig.\ref{Figsh} and the inset. The only fitting parameter was the activation energy  $U_c$. All other quantities in Eqs.(\ref{I},\ref{V}) were determined from experiments. We can see from figure that calculations describe the experiments well.

The noise in the voltage region   $V_d < V < V_s$ is related to the expectation of a quite large fluctuation transforming the electron system from one local energy minimum to another. The intense noise in the nonlinear regime and the two-threshold flow disappear earlier than the MIT is reached \cite{miriam}. Such a behavior agrees well with the noise measurements in the linear regime, where the saturation of  $\alpha$  at the level     $\alpha \simeq 2$, corresponding to the usual amorphous phase was observed at an electron density noticeably lower than  $n_c$.

\par\bigskip

{\bf 2.6.2. Magnetic properties of the insulating phase}

\par\bigskip

{\it Localized droplets.} Experimental data presented in \cite{teneh} were interpreted as the result of the existence of localized droplets in the insulating phase, \textit{i.e.}, localized formations resembling quantum dots consisting of a few ($\simeq 4$) electrons. Electrons in such droplets are completely spin-polarized with a random orientation of the total magnetic moment in the zero magnetic field.

\begin{figure}
\scalebox{0.8}{\includegraphics[width=\columnwidth]{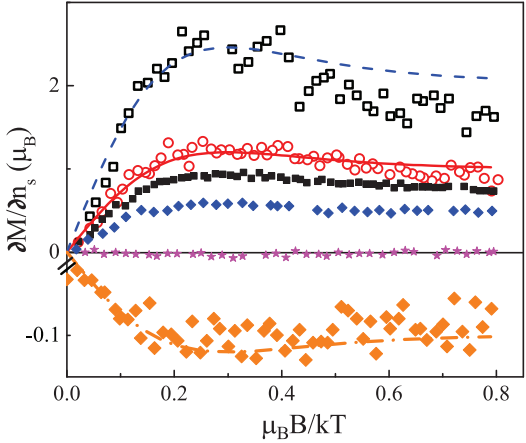}}
\caption{Derivative of the magnetic moment of a unit area by the electron density as a function of the normalized magnetic field parallel to the MOSFET   interface.   Unfilled   symbols:    $n_s=5\times10^{10}$ cm$^{-2}$: squares  $T=0.8K$, circles $T=1.2 K$. Filled symbols $T=1.8$~K:  squares,  diamonds, stars, and large diamonds correspond to respective electron densities   $0.4, 0.8, 1.4, 2.5 \times 10^{11}$ cm$^{-2}$ correspondingly. The  dashed,  solid,  and dashed-dotted curves are fittings using Eq.(\ref{der}). From \cite{teneh}. }
\label{Figten}
\end{figure}

Despite identical names, it is unlikely that localized droplets have something in common with the free droplets proposed in  \cite{kiv}
as one of the intermediate phases between metal and Wigner crystal.

In the framework of the concept of localized droplets, the magnetic moment of a unit area can be written as
\begin{equation}
M=  \mu _B [n_d \tanh(sb)+ (n_s-n_d)\tanh(b)],
\label{mag}
\end{equation}
where $n_d$ is the electron density in droplets, $s$ is the mean number of electrons in one droplet, and 
$b=\mu_b B/kT$ is the normalized magnetic field. For simplicity, we assume that $s\gg 1$, $b \ll 1$. Then
\begin{equation}
\frac {\partial M}{\partial n_s}  \simeq N_d \frac {\partial s}{\partial {n_s}} [ \tanh(sb) +sb \cosh^{-2}(sb) ],
\label{der}
\end{equation}
where $N_d =n_d/s$ is the number of droplets (strong pinning centers), weakly depending on the electron density in the insulating phase but depending on temperature.

Fig.\ref{Figten} shows fitting curves based on Eq.(\ref{der}) for the  values  of  the  fitting  parameter 
$ N_d \frac {\partial s}{\partial {n_s}}$ equal to  2, 1, -0.1, and $s=4$. We can see that the fitting curves describe the
experiments well.

A comparison of calculations with experiments leads to some interesting conclusions. First, the value of the derivative $ \frac {\partial M}{\partial n_s} \simeq2$ at  temperature   $T=0.8$~K means  that  at lower densities, this derivative is considerably smaller than unity and the interval of its large values is quite narrow. This statement is inconsistent with higher-temperature measurements (see, \textit{e.g.}, Fig.\ref{Figpud1}.) Second, the number of localized droplets turns out to be temperature-dependent (which requires additional verification, however) and weakly dependent on the electron density in the insulating phase. Third, after transition to the metallic phase, the derivative  $\frac {\partial s}{\partial {n_s}}$ changes sign. 

A question naturally arises: How are the concepts of localized droplets and the amorphous phase following from noise measurements described above related? Taking into account that the Coulomb energy considerably exceeds the temperature for  $T \leq 2$~K, we see that the characteristic spatial scale between electrons in a droplet should not differ significantly from the mean distance between electrons. In other words, the density of an electron system with droplets weakly changes at scales exceeding (slightly) the mean distance between electrons, which corresponds to the amorphous state and is qualitatively confirmed by experiments, albeit those performed with samples of different qualities prepared by different manufacturers.

{\it Magnetization and spin susceptibility in the insulating phase.} We again consider Fig.\ref{Figpud1}.  In a magnetic field of 2~T, the magnetic moment increases linearly up to a density of $5\times10^{10}$ cm$^{-2}$. As the electron density is increased further, the magnetic moment continues to increase, but now proportionally to  $n_s$. Because at $B=2$ T  the temperature 1.7~K corresponds to the saturation of the magnetic moment of droplets, we have to conclude that for $n_s>5\times10^{10}$ cm$^{-2}$, not  all  localized  electrons  enter  droplets. Therefore, the number of strong pinning centers is  $n_d \sim 5\times10^{10}/s \sim 1.2\times10^{10}$ cm$^{-2}$ ($s=4$), and in the metallic phase, according to the estimate in Section 2.3.3,  $s \sim2$.

In  the  magnetic  field   $B=5$~T at  the  temperature    $T=0.4$~K, the parameter  $b \simeq 16$, and all electrons in the insulating phase are spin-polarized. It can be expected that the straight line $M=\mu_B n_s$ specifies the behavior of the magnetic moment in the insulating phase, in agreement with the results in  \cite{rez,ten}. However, as follows from Fig.\ref{FigM}, experimental points in the metallic phase at the minimal density are higher than this straight line, which is unsurprising because the Land\'e factor is $g>2$ in the metallic phase.

 The susceptibility in the insulating phase is determined by the initial region of the curve describing the dependence of the magnetic moment on the magnetic field. In the concept of droplets
 \begin{equation}
 \chi= \frac{\mu_B^2}{kT}[n_d(s-1)+n_s].
 \label{chi1}
 \end{equation}
In Ref.\cite{teneh} the relation  $n_d \propto 1/T$ was observed to roughly hold. 

In the initial part of the dependence  $\chi(n_s)$, the susceptibility is proportional to density (see Fig.\ref{Figpud2}); ); therefore, $n_d=n_s$. The slope of the initial part of the sample shown by squares in Fig.\ref{Figpud2} is 30\% smaller than that for a sample whose data are shown by circles. According to Eq.(\ref{chi1}), this means that the mean number s of particles in a droplet is smaller for a more perfect sample than for a more disordered sample.

\par\bigskip

{\bf 2.7	Additional intermediate conclusions}\\

We note that neither of the two alternative naive models presented in Section 2.3.1 is fully correct. Only their combination is consistent with experiments.

 Indeed, in the metallic phase, both the energy-averaged effective electron mass  ($m^*$) and the effective electron mass at the Fermi level ($m^*_F$) increase with decreasing the electron density.  We  note  that  $m^*_F$ exhibits  a  tendency  to diverge, in agreement with the second model, assuming  that
localized electrons are absent in the metallic phase and the properties of the electron system are determined only by interaction. At the same time, the metallic phase undoubtedly contains a small number of localized electrons, which, for example, determine the magnetic susceptibility of the electron system. The number of localized electrons depends on the sample quality. Experiments demonstrate that according to the assumptions of the first model, the presence of localized electrons in the metallic phase does not affect the Hall effect in
weak fields.
 
 The insulating phase in the general case contains electrons included in localized droplets, with the total magnetic moment of the droplet behaving as a single whole, and also localized electrons not entering the droplets. The relation between these two electron groups probably depends on the sample quality.  Some properties of this mixture, for example, noise characteristics, are similar to those of the amorphous phase. Unfortunately, any theoretical calculations concerning the magnetic properties of a strongly coupled disordered electron system are absent. 
Studying the magnetic moment at the MIT at millikelvin temperatures could be of great interest. The extrapolation of the available information predicts a jump in the magnetic moment in this region.

\par\bigskip

{\bf 3.  Electrons in SiGe/Si/SiGe quantum wells}

\par\bigskip

{\bf 3.1. Advantages and disadvantages of the structures}\\

The properties of electrons in the highest-mobility SiGe/Si/ SiGe quantum wells were studied in Refs.\cite{lu,lu2,lu3,mel2,mel,mel1}. We consider the results of studies \cite{mel2,mel,mel1}, in which electrons were located in a (100) silicon quantum well  $150$ \AA wide. The quantum well is bounded from above and at the bottom by SiGe barriers.  The top barrier $\sim 1500$\AA wide was covered with a $10$\AA silicon  layer  with  thermally  deposited ($2000-3000$ \AA  thick) SiO layer and a metallic gate. The samples, as in the Si-MOSFET case, had the shape of Hall bridges.

\begin{figure}
\scalebox{0.8}{\includegraphics[width=\columnwidth]{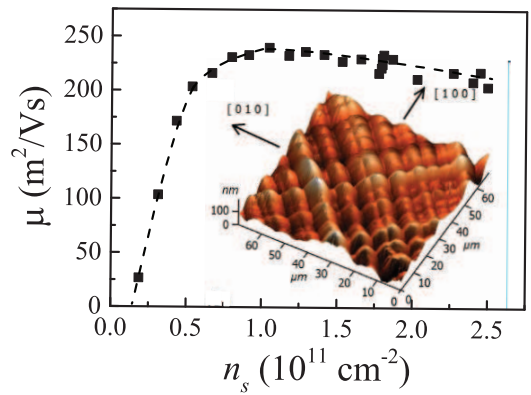}}
\caption{Electron mobility in a (100) SiGe/Si/SiGe quantum well as a function of the electron density at $T=50$ mK. The inset shows an atomic-force microscope scan of the sample surface (before the deposition of an SiO insulator and a metallic gate)  From \cite{mel2,mel}. }
\label{FigSiGe}
\end{figure}

The advantages of the electron system in quantum wells in SiGe/Si/SiGe systems are due to, first, the high electron mobility and, second, the feasibility of measurements at low electron densities. The dependence of electron mobility on the electron density for one of the best samples is shown in Fig.\ref{FigSiGe}. We can see that the maximum electron mobility in a SiGe/Si/SiGe quantum well is almost two orders of magnitude greater than the electron mobility in the best Si-MOSFETs. In addition, metallic conductivity is observed down to very low electron densities of the order of  $1.5\times10^{10}$ cm$^{-2}$. 

Among the disadvantages is a weaker (at a fixed density) electron-electron interaction. As already mentioned in the Introduction, the reason is the dielectric constant greater by a factor of 1.5 and a greater spread of the electron wave function in the direction normal to the interface.

Another disadvantage of SiGe/Si/SiGe structures is a complex surface relief (see the inset in Fig.\ref{FigSiGe}). The relief has ``ridges'' extended along the [110] and [-110] directions with a characteristic height of  $\sim 60$~nm and period of  $1\mu$m. The relief is rather flat because the period greatly exceeds the characteristic ridge height. It was shown in  \cite{mel} that the potential well relief repeats the surface relief, and therefore the modulation of the electron density caused by ridges is virtually absent.

Nevertheless, the relief modulation considerably complicates measurements in a magnetic field parallel to the surface. Although the field is parallel to the surface on average, the bends of the quantum well lead to the appearance of a local normal (in the ideal case, alternating) component. We can no longer assume that the parallel field acts only on the electron spin, because the local normal component acts on the orbital motion. This difficulty can be eliminated by a proper choice of the orientation of the Hall bridge with respect to crystallographic axes and the magnetic field orientation with respect to the measuring current \cite{mel,mel1}.

\par\bigskip

{\bf 3.2. Tendency of a flat band to appear at the Fermi level}\\

\begin{figure}
\scalebox{0.8}{\includegraphics[width=\columnwidth]{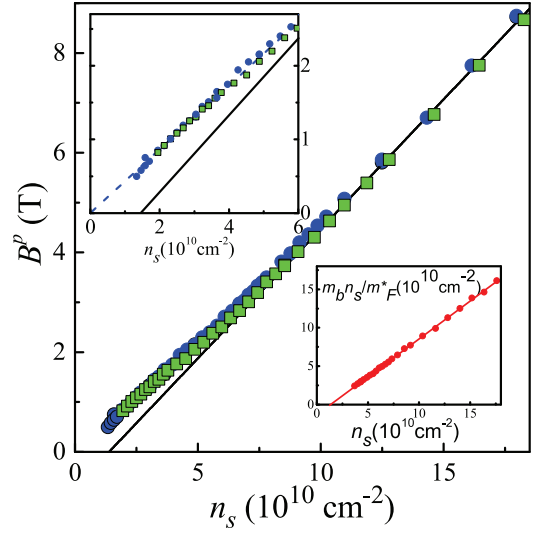}}
\caption{Total spin-polarization fields for two samples (different symbols) at temperature 30 mK as a function of the electron density. The straight line is a linear fitting to points at large densities. The left inset is the initial part of the same dependence on an expanded scale. The right inset is the density dependence of the inverse effective mass at the Fermi level \cite{mel1}. }
\label{Figmish}
\end{figure}

The electron system in SiGe/Si/SiGe quantum wells was used for measurements of two types \cite{mel1}. First, the total spin- polarization field was measured as a function of the electron density. The corresponding results are shown in Fig.\ref{Figmish}.

The observed behavior of the total spin-polarization field is  consistent (albeit up to a numerical coefficient) with Monte Carlo simulations \cite{waintal}. At high densities, the dependence is well approximated by a straight line tending to a finite density as $B^p \rightarrow  0$. At a density approximately equal to  $n_s \simeq 7\times10^{10}$ cm$^{-2}$, the dependence exhibits a break and  $B^p(n_s)$ tends to the origin at lower densities (the dashed line in the left inset in Fig.\ref{Figmish}).

The right inset in Fig.\ref{Figmish} shows the dependence of  $m_b n_s/m^*_F$ on $n_s$. We  discuss  the  method  for  experimentally measuring the electron mass at the Fermi level below, and now note the linear dependence in the right inset extrapolating to a finite electron density, which coincides, within the experimental accuracy, with the result of the extrapolation of the dependence 
 $B^p(n_s)$. As mentioned above, this dependence is determined by the energy-averaged mass. Therefore, at high densities, the energy-averaged electron mass and the electron mass at the Fermi level are at least proportional to each other, if not coincident. At the same time, at densities $n_s < 7\times10^{10} c m^{-2}$ the behavior of the masses is completely different: the electron mass at the Fermi level continues to increase with decreasing electron density, whereas the energy-averaged mass saturates.

\begin{figure}
\scalebox{0.75}{\includegraphics[width=\columnwidth]{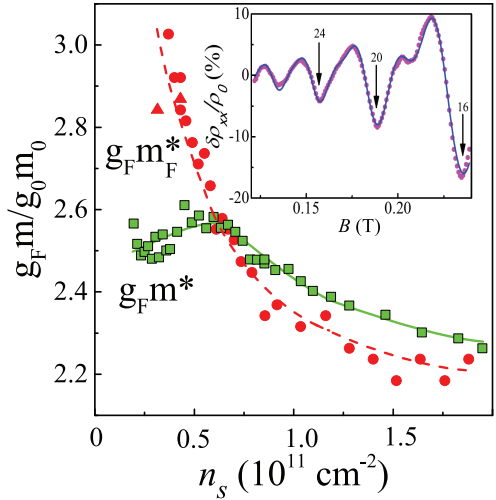}}
\caption{Comparison of the density dependences of the energy-averaged electron mass and the mass at the Fermi level. $T=30$ mK. The lines are guides to the eye. The inset illustrates the quality of fitting the experimental dependence of the normalized resistance using equation (\ref{A}). From \cite{mel1}. }
\label{Figmisha}
\end{figure}

Such a behavior is more clearly demonstrated in Fig.\ref{Figmisha}, where  the  dependences  of   $g_F m^*$ and $g_F m^*_F$ on  the  electron density are compared. The first dependence was found by using experimental points  $B^p(n_s)$ and the relation
\begin{equation}
g_F\mu_B B^p=\frac{2\pi\hbar^2n_s}{m^*g_v},\label{gm}
\end{equation}
where $g_v=2$ is the valley degeneracy. The dependence $g_Fm^*_F (n_s)$ was obtained from Eqs.(\ref{A}). The dependence of the resistance on the magnetic field was fitted as shown in the inset in Fig.\ref{Figmisha}. The fitting parameters were  $m^*_F$, $T_Dm^*_F$,  and  $g_Fm^*_F$. The  value  of $m^*_F$ was  found  in  separate  experiments	  from  the  temperature  dependence  of  quantum oscillations with an accuracy of 10\%. The fitting of experimental curves turned out to be not very sensitive to the parameters 
 $m^*_F$ and $T_Dm^*_F$, but was rather sensitive to the product    $g_Fm^*_F$. 

\begin{figure}
\scalebox{0.7}{\includegraphics[width=\columnwidth]{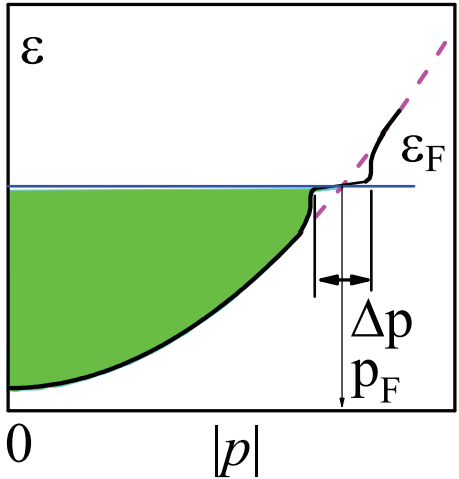}}
\caption{ Assumed single-particle spectrum with a flat region at the Fermi level. The energy and momentum region occupied by electrons is shown by filling. }
\label{Figkhodel}
\end{figure}

Comparison  of  the  behavior  of   $g_Fm^*_F$ and $g_Fm^*$ at densities $n_s <7*10^{10}$ cm$^{-2}$ clearly  demonstrates  the
difference: the electron mass at the Fermi level continues to increase with decreasing density, whereas the energy-averaged mass saturates. Such a behavior corresponds to the appearance of a flat band in the single-particle spectrum at the Fermi level (see Fig.\ref{Figkhodel}).

In  the  vicinity   of   the critical  density $\simeq 1.4*10^{10}$ cm$^{-2}$,  the electron system is in the critical region where the effective mass at the Fermi level is limited by the
temperature  $m^*_F < p_F \Delta p/4kT$.  The data presented in Fig.\ref{Figmisha} 
give the estimate of the interval  $\Delta p$: $\frac{\Delta p}{p_F} =0.06$.

We note that the appearance of the interaction-induced flat band of the single-particle spectrum at the Fermi level was predicted in theoretical papers (see, \textit{e.g.}, Refs.\cite{khod, kotl}) based on an absolutely different approach.

Both experiments presented above and the conclusion on the different behaviors of the electron mass at the Fermi level and the energy-averaged mass are quite unusual. To confirm this behavior, independent experiments are required. Such an experiment was discussed in 
 \cite{dol18}.

\par\bigskip

{\bf 4.   Conclusions}\\

By comparing the results obtained by different research groups for different Si-MOSFET samples with the results of measurements with SiGe/Si/SiGe quantum wells, we can conclude  that  the  effective  electron  mass   $m^*_F$ at  the  Fermi level  in  the  metallic  phase  tends  to  diverge  as  the electron density decreases. The results of measurements of  the effective mass at the Fermi level performed by different groups with samples of the same type from different manufacturers differ only due to different ways of the data analysis. Unfortunately,  the  increase  in  $m^*_F$ in  the  region  available  for measurements is not as considerable as, for example, in $^3$He \cite{he,he1}, and the conclusion about the divergence of the mass has to be made based on extrapolation.

In the metallic phase, a small amount (depending on the sample) of localized electrons weakly affecting transport properties is observed.

In the insulating phase and in the MIT vicinity, the electron system reveals properties typical of amorphous media with strongly interacting particles. The study of the microscopic structure of the insulating phase in Si-MOSFET has shown that it consists of localized droplets (resembling quantum dots) containing on average about four spin-polarized electrons, and of localized electrons outside such droplets.

A simple listing demonstrates the considerable recent progress in experiments that has been achieved due to the development of experimental methods. Unfortunately, the quality of Si-MOSFETs has not improved. On the other hand, the study of electrons in SiGe/Si/SiGe quantum wells is far from comprehensive, and conclusions made based on the available experimental data are only preliminary. Independent additional experiments are described in \cite{dol18}. Nevertheless, to reliably prove the possible independence of the MIT from events at the Fermi level, further experiments are required.

\par\bigskip

\textbf{Acknowledgments}\\

The author thanks A.~A. Shashkin and S.~V. Kravchenko for  useful discussions. The author is especially grateful to  V. M. Pudalov for his reading of the manuscript and giving numerous remarks. The study was partially supported by the Russian Foundation for Basic Research (grant no. 18-02-00368) and the State Program of the Institute of Solid State Physics, RAS.

\end{document}